# Cyber-Physical War Gaming

EJM Colbert, DT Sullivan, A Kott

*Computational and Information Sciences Directorate*
*U.S. Army Research Laboratory, Adelphi, MD, USA*
E-mail: edward.j.colbert2.civ@mail.mil; daniel.t.sullivan12.ctr@mail.mil;
alexander.kott1.civ@mail.mil

**Abstract:** *This paper presents general strategies for cyber war gaming of Cyber-Physical Systems (CPSs) that are used for cyber security research at the U.S. Army Research Laboratory (ARL). Since Supervisory Control and Data Acquisition (SCADA) and other CPSs are operational systems, it is difficult or impossible to perform security experiments on actual systems. The authors describe how table-top strategy sessions and realistic, live CPS war games are conducted at ARL. They also discuss how the recorded actions of the war game activity can be used to test and validate cyber-defence models, such as game-theoretic security models.*

## Introduction

It is commonly known that many Cyber-Physical Systems (CPSs)—which include Industrial Control Systems (ICSs), Supervisory Control and Data Acquisition (SCADA) systems, and "things" in the so-called Internet of Things (IOT)—are often vulnerable to cyber threats. For example, the very well known STUXNET attack targeted nuclear power plant centrifuges in Iran. Similar cyber (and physical) attacks on CPS equipment could very well target the United States' power grid, the control systems of large water dams, and even people, since wearable CPSs are becoming more and more popular. Even though the hardware controlling CPSs are vulnerable to cyber-attacks, they remain relatively unprotected from many potential adversaries. In an attempt to educate operators, owners and users of CPSs, the ARL recruits players for red (attack) and blue (defence) teams and conducts CPS war games in an environment as close to realistic as possible. The decisions (moves) of the war game players are recorded so that a better understanding of how security models may allow useful assessments and/or predictions of attacks against currently operating SCADA and other CPS systems. This paper discusses current results and future plans for CPS war gaming in the following subsections.

## SCADA and Industrial Control Systems

An Industrial Control System (ICS) is a generic term for a combined system of electrical and mechanical devices and processes that automatically control the operation or one or more physical machines. Machines can be as simple as a self-service gasoline filling pump system, or as complex as a complete robotic assembly line in a vehicle production plant.

A Supervisory Control and Data Acquisition (SCADA) system is a specific type of ICS that usually controls many machines separated by relatively large distances (Sullivan, Luiff & Colbert 2016). The key discriminator which separates an ICS from an Information Technology (IT) system is that an ICS monitors or interacts with something physical in the real world. Previously, ICSs were isolated; however, due to demand for greater productivity and efficiency, ICSs and IT enterprise networks are now being inter-connected. This new trend exposes ICS devices to many new threats they would have never encountered when previously isolated.

**Error! Reference source not found.**, below, depicts a typical manufacturing plant with its ICS connected to a corporate IT network. The ICS consists of the Supervisory Control and Basic Control layers and field bus networks connected to plant sensors and other physical devices. In this example, a corporate Demilitarized Zone (DMZ) on the edge of the corporate network offers some protection against cyber-attacks from external networks. However, many network attack vectors remain, putting the physical machines (sensors and equipment in the figure) at risk. In many cases, physical machines in ICS and SCADA systems are critically important for human sustenance, such as power switches and transformers in the US electric grid. Automated processes in the ICS are programmed into and controlled by ICS hardware devices such as Programmable Logic Controllers (PLCs) and are consequently monitored by a human operator, who manages Human Machine Interfaces (HMIs). A recent summary of cyber security issues and method of SCADA and ICS systems was written by Colbert and Kott (2016).

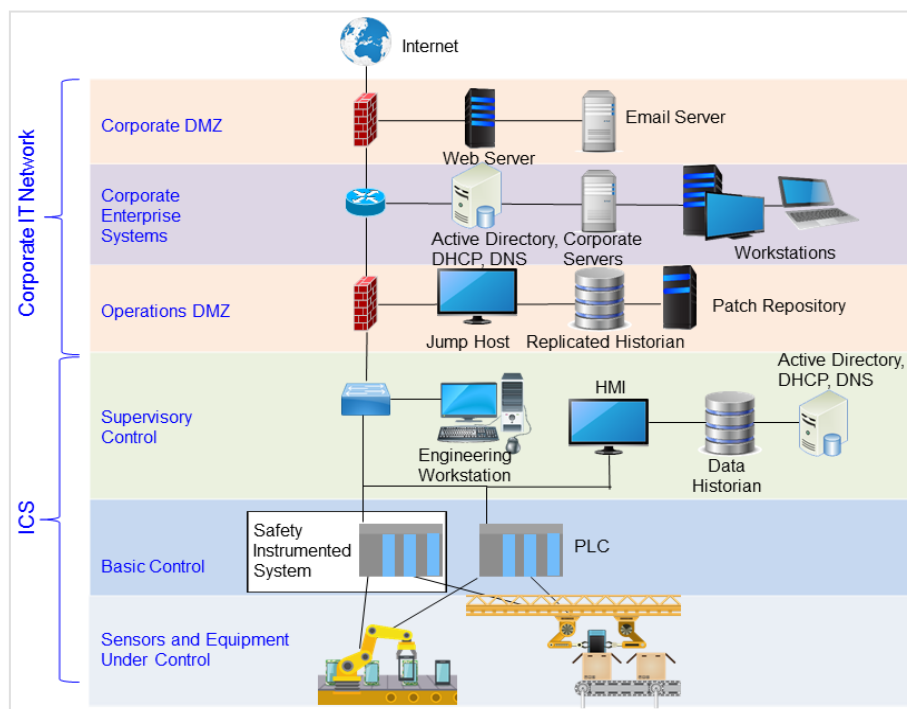

**Figure 1:** Notional ICS and corporate IT network architecture

## War Gaming

War gaming has been used to refine aspects of warfare since the 5th century B.C. when the ancient Chinese general and philosopher Sun Tzu documented military strategies in *The art of war* (Sawyer & Sawyer 2007). War gaming methods are now used in business and military training as essential elements of developing strategic leaders (Horn 2011; Curry & Perla 2011; McCown 2005). Without risking a company's reputation, stock value, or customers' confidence, corporate cyber defenders can test their cyber play books and identify needed skills or new processes in a safe and controlled environment. A realistic war game can be an immersive learning tool where players can practice making decisions in real time and can see the effects immediately. Modern cyber war games are played between cyber attacker in a 'red team' and cyber defenders in a 'blue team' and are organised around a business scenario in which teams and players receive awards for their actions. They are necessarily structured to simulate a real attack so that the defenders can exercise and refine their defence strategy methods for potential future engagements (for example, Bailey, Kaplan & Weinberg 2012). War games are also very useful for uncovering 'black swans' or unforeseen gaps in current cyber defences (Perla & McGrady 2011). Knowledge learned in war games is dependent upon the ability of the game moderators to simulate real-world network and processes.

## Table-Top War Gaming at ARL

In order to develop strategies for defending SCADA systems, a full-day Table-Top exercise was performed at ARL in 2014. A SCADA system named AQUA was analysed during the exercise. AQUA is a notional Army SCADA system representative of an Army production system. The process diagram for AQUA is shown in Error! Reference source not found., below. AQUA produces high-quality meals for soldiers and consists of six manufacturing processes indicated in the figure.

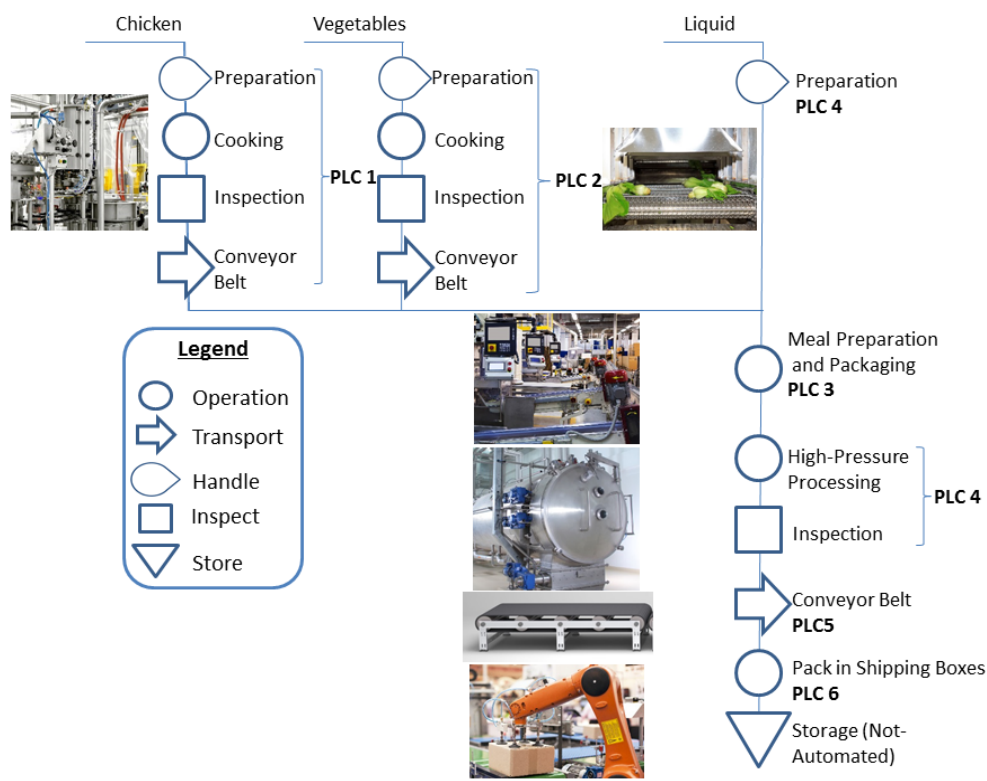

**Figure 2:** Process map of the AQUA ICS

The plant network diagram for AQUA is shown in Error! Reference source not found., below. It consists of 6 PLCs, several Windows workstations, a Closed-Circuit Television (CCTV) system, and a wireless network for tablet computers. Technicians use the tablet computers to access the Human Machine Interface (HMI) displays using Hypertext Transfer Protocol Secure (HTTPS) and to read machinery documentation in Portable Document Format (PDF). The plant network is not connected to the corporate network or the Internet. The HMIs and PLCs transmit and receive the commonly used ModBus-transmission control protocol (TCP) communication protocol. The AQUA system is air-gapped from the corporate network and the Internet, as indicated in the figure.

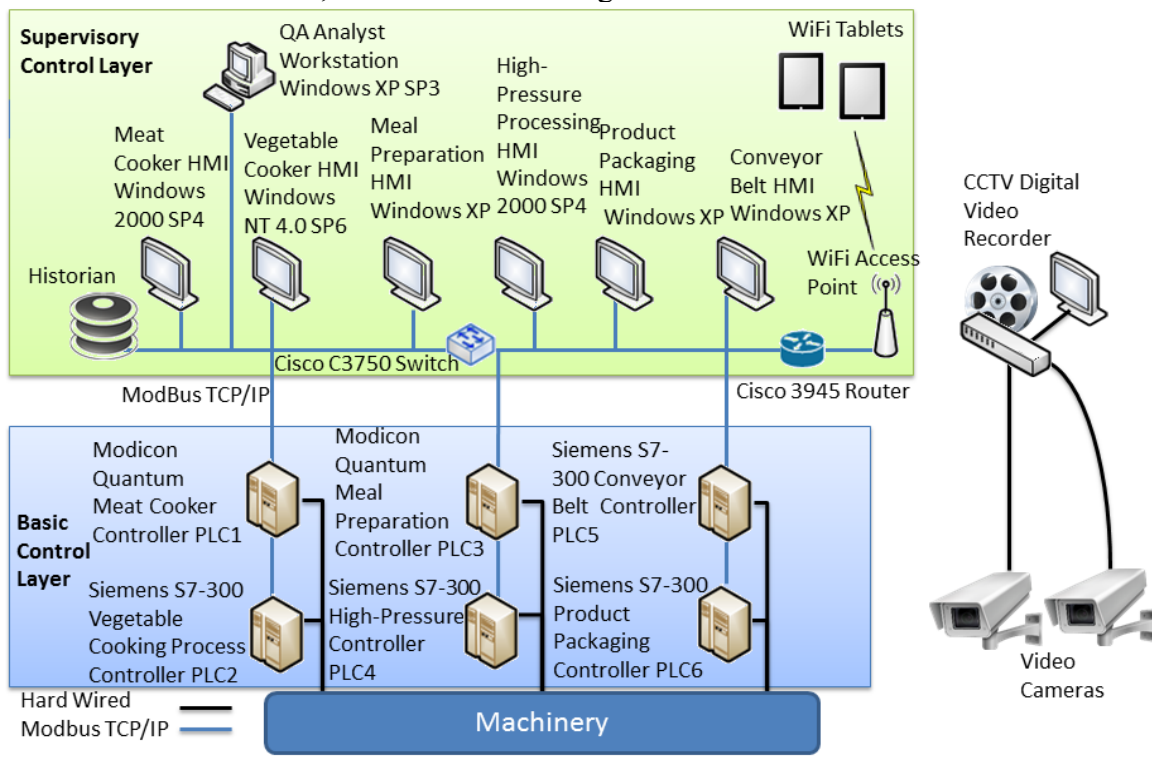

**Figure 3:** Network diagram of the AQUA plant network

The Table-Top activity was conducted in a war gaming scenario in which a red team would formulate attacks, and a blue team would attempt to mitigate those attacks. The teams were composed of government and contract workers from ARL. In general, the network-based attacks proposed by the red team were defendable, often by simple security measures that were unfortunately not already installed in the AQUA system. Some specific areas of follow-up research and development were proposed for SCADA intrusion detection, such as a Layer-2 Intrusion Detection System (IDS) and anomaly-based network detectors.

An information packet (Colbert *et al.* 2015a) was given to all red and blue team participants before the exercise began. It provided background material for those not familiar with SCADA systems, a discussion of the differences between IT enterprise networks and SCADA networks. Technical information about the AQUA system was also provided in the packet.

The general goal of the Table-Top exercise was to generate enough discussion so that general gaps in security measures could be identified. The ARL Sustaining Base Network Assurance Branch (SBNAB) and Network Security Branch (NSB) support research and development activities of IT enterprise IDS systems and most of the game participants from these branches were familiar with those IDS efforts. As such, improvements to SCADA security were typically framed in terms of IDS plug-in modules. Complete details of the exercise are documented in Colbert *et al.* (2015b). Some general conclusions from the exercise follow.

- The SCADA network should be regarded in the same manner that a government classified network is regarded. While the system may not be classified, the same

- measures should be applied: tight physical security; zero or highly restricted wireless access; US Department of Defense (DoD) Security Technical Implementation Guidance (STIG) regulations should be applied when possible; firewalls should be in place to highly restrict traffic to and from critical network assets; and, appropriate policies and training should be created and enforced.
- Two particular technology improvements were noted for follow-up research and development for IDS plug-in modules:
    - Anomaly-based network monitoring should be developed for HMI/PLC traffic.
    - Layer-2 traffic should be monitored for defending against attacks coming in below the Internet Protocol (IP) layer (Layer-3). This includes wireless traffic, which seems to be coming to more and more 'protected' networks, even classified government networks.
- A considerable amount of development effort would be needed to develop new security capabilities for SCADA, and a considerable amount of maintenance effort would be needed by both network security analysts and plant operators to keep the new security tools calibrated, especially after changes are made in the plant network. A benefit of having a SCADA anomaly detector in place, however, is that all changes to the plant network would be immediately alarmed.
- Since much SCADA hardware is not designed for cyber security, physical security often currently provides the main defence. Defence-in-depth is vitally important for an effective defence. Strong policies and employee training must be used, especially for alleviating insider threat.

The 2014 Table-Top exercise provided an exceptional view of the technical and programmatic security issues in SCADA systems that could be followed up by hardware test bed experiments. ARL plans to continue to perform similar exercises in the future to better understand how ARL SCADA research should proceed and to better educate its personnel about ongoing problems and issues specific to different ICSs.

It is important to note that while important general conclusions for improved security were found in the exercise, since each SCADA system is unique a dedicated Table-Top or other brainstorming activity would be needed if threats are to be identified and mitigated for specific systems. For the AQUA production line, poor network security proved to be a severe problem. Poor network security exists on many modern SCADA systems since they were originally designed for physical security protection in an isolated network environment. Consequently, most or all of the red team attacks from the Table-Top experiment would have been successful. However, a few simple network security improvements would have made the AQUA system significantly safer, such as removing the wireless access point and not using the default configuration for routers and switches.

In general, pure network attacks against AQUA are defendable if the IT network section of the system is upgraded with appropriate security measures. However, there are non-IT network elements of the AQUA system that are more complicated to defend; PLCs are not built for network security and the Modbus TCP protocol does not use any authentication measures.

Since SCADA systems are, overall, distinctly different from standard IT enterprise systems, network security analysts need to educate themselves and understand the intricacies of

SCADA in order to guard them properly. Furthermore, for each SCADA installation, the analysts would need to understand the *particular* intricacies of the SCADA system being defended, since each SCADA system usually has very different types of hardware. This differs from IT enterprise systems, which typically run a limited number of operating systems and applications. Unlike SCADA systems, the underlying hardware in IT enterprise systems is much less relevant, since many or most of the vulnerabilities are software-oriented and already known.

**Live war gaming at ARL**
The objective of the war gaming activity on a real live SCADA system is to capture activities of cyber defenders as they protect a corporate and an ICS network as well as capture the actions of hackers as they try to halt automated processes. Afterwards, the captured activities are compared to a security model of the system to determine if the model algorithm can accurately predict actions of the cyber actors.

In the planned live war game, a simulated corporate network and a SCADA system are monitored by a security operations centre while being attacked by hackers. Three teams of actors participate in the war game:

- **Red Team:** threat actors who represent a mid-level of knowledge in penetrating networks. Red team members are not 'script-kiddies' nor do they have the resources of a nation-state;
- **Blue Team:** security operations analysts who will monitor and defend the corporate network and the SCADA system operations; and,
- **White Team:** neutral group that facilitates the war game. The white team provides the initial configuration of the corporate network and SCADA system, training (as needed), and monitors all activity of the game. The white team also adjudicates scoring of the red and blue teams.

The mission of the red team is to compromise the availability of the SCADA system operation by stopping the automated process. The mission of the blue team is to monitor system status, maintain critical IT services in the corporate network, and maintain normal operations of the plant. At the conclusion of each war game event, the white team will capture all artefacts and archive the data for analysis. A post-war game meeting will be held to review the actions of each team and share lessons learned.

**Sample live war game scenario**
A specific sample live war game is described next, in which two companies manufacture competing smartphone products. Company ALPHA has a SCADA production line that is connecting to its corporate network and to the Internet. The ALPHA corporate network and smartphone production line system are protected by the blue team in the war game scenario. Company BETA has recently suffered a drop in smartphone sales and wishes to increase its revenue and dominate the market. Company BETA hires a team of hackers (the red team) to compromise the availability of the company ALPHA smartphone production line. Sample game play (schedule, rules, and point scoring system) for this war game are described next.

**War game schedule**
The blue team must design its network defences prior to the start of the war game. The white team will provide a read-ahead package to the blue team members one-to-two weeks before

the war game begins. The read-ahead package will contain technical information about the system, such as:

- Initial network topology of the simulated corporate and ICS network for the war game;
- Lists of each physical and Virtual Machine (VM) used in the war game along with their respective Operating System (OS) and software;
- Make and model of the PLC(s) which operate the simulated automated process; and,
- Additional hardware such as firewalls, switches, or routers that are available to the blue team for enhancing the system security.

The blue team members may bring additional tools or software they would like to install. Red team members may also bring their own tools and software. Red team members will not receive the read-ahead package. They will only receive general information about the war game play provided by the white team in the kickoff meeting.

The war game will be executed in two events (Event 1 and Event 2), which are identical except for the fact that the blue team has additional SCADA security training in Event 2. There will be two distinct blue teams (Blue Team #1 and Blue Team #2) as well as two distinct red teams (Red Team #1 and Red Team #2). Each event will be conducted over a 3-day period. The white team will meet with the blue and red teams in advance of the war game to answer questions and coordinate building access.

The timeline for Event 1 is illustrated in **Figure 4**, below. On Day 1, both teams will meet for a short introductory kickoff meeting. Blue Team #1 will then have two hours to install software or hardware to protect the simulated corporate and ICS networks. An additional hour will be spent with the white team discussing and verifying the mitigations. The white team will then commence the war game and the blue team will monitor and attempt to defend the ALPHA SCADA system for the next 21 hours while the red team attempts to compromise its availability. During game play, activities of each player as well as network packet data will be captured by the white team. Each red and blue team member will keep notes on the actions taken to validate point scoring.

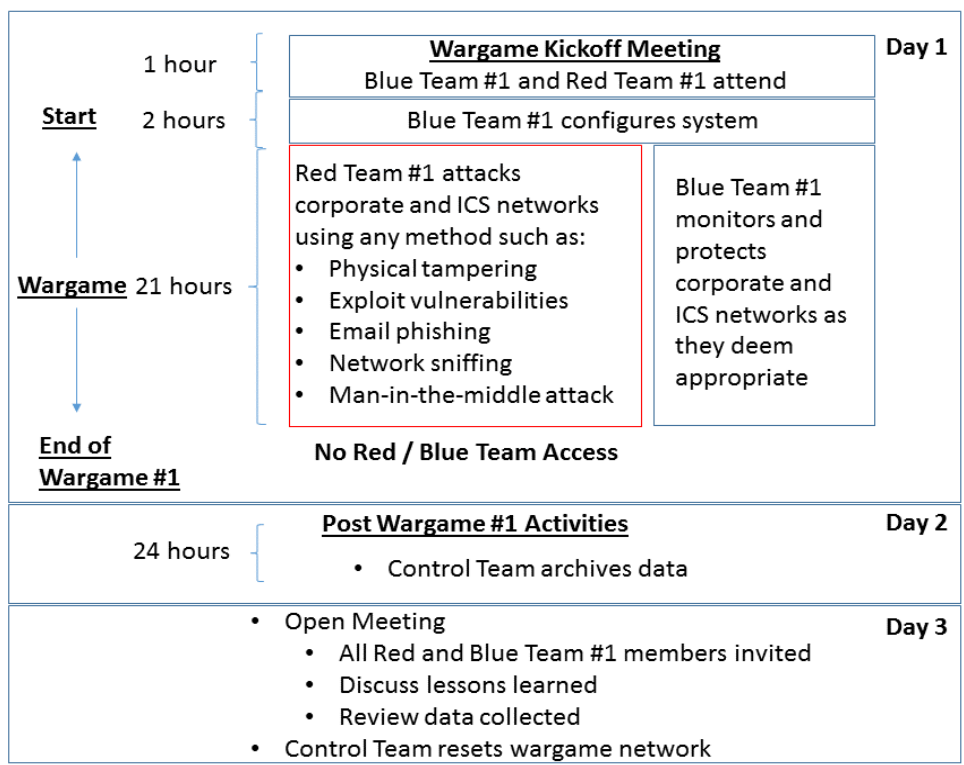

**Figure 4:** Event 1 timeline

After 24 hours of game play (Day 2), the white team will stop the game play for Event 1. Each player will provide notes describing his or her actions to the white team. The white team will then archive game play data such as network packet captures, logs, and capture of screen video. The white team will have 24 hours to complete the data capture and archiving of the event.

On Day 3, all red and blue team members will be invited to a two-hour technical exchange meeting during which each team will share insights and information about their strategies and provide feedback to the white team. Point totals will be share and game prizes will be awarded. The white team will brief their analysis of the collected game play data and will discuss conclusions of the war game. Lessons learned will be recorded by the white team.

## Event 2

Event 2 is conducted in a similar manner to Event 1 with one change. Before starting the event (on Day 4), the white team will provide an in-depth security brief on general methods of SCADA security and specific applications for the ALPHA production-line system. This change is performed in order to evaluate the effectiveness of blue team SCADA security training. Many cyber defence security workers are well trained in IT enterprise security, but not necessarily in SCADA security. The three-day structure of Event 1 shown in **Figure 4** is then repeated for Event 2 during days 4, 5, and 6, respectively. A new set of red and blue teams are chosen using the same team selection criteria used for Event 1.

## Sample war game play

The red and blue teams are provided with game rules and objectives used for scoring in the initial kickoff meeting. Violations of game rules create point penalties for the respective teams. A preliminary set of rules and a sample point scoring system are given below.

## General game structure

In this example exercise, the white team randomly selects members of the red and blue teams from a larger pool of volunteers. Players in event 1 are not allowed to participate in event 2. Members from the red and blue teams will nominate their own team captain. In order to accomplish their strategy, red and blue team members are allowed to use their own hardware and software during the game activity. The red team may use cyber-attacks or physical attacks as they see fit, since both are viable in a realistic setting. Game play is dictated to the teams using specific rules of conduct and point scoring, such as those listed in **Table 1**, below.

| Game Rule | Description | Violation Penalty Points |
|---|---|---|
| 1 | A competitor can participate in only one war game event | 5 points deducted from the team for each person who attempts to compete in both war game events |
| 2 | Before competition begins, the red team may not eavesdrop or surreptitiously observe the blue team configure the war game network | 20 points deducted from the red team |
| 3 | During competition, red and blue team members may not visit their opponents' area unless approved by the white team | 20 points deducted from the offending team |

**Table 1:** Red and blue team game rules of conduct

## Sample Point Scoring

The white team will track game scores. Negative points are levied for violations of game rules and positive points are awarded for achievements. In this sample game, each team has knowledge of its current score, but it does not have knowledge of the other team's score. **Table 2**, below, lists sample point values for each achievement scored by the blue team, as well as the criterion for award and the award validation method.

| Achievement | Point Value | Award Criterion | Validation Method |
|---|---|---|---|
| 1 | 5 points for each critical service per hour | Critical service is normally running | White team will connect to each service to verify it is operational |
| 2 | 10 points | Stopping red team intrusion | Blue team shows the white team that an intruder was present and the blue team's action prevented new connections from the intruder. The blue team must show that the intruder is being prevented from reconnecting by showing log entries to the white team. |

**Table 2:** Blue team point scoring system

The red team gains points for having presence on the corporate network or the ALPHA SCADA system, disabling vital ALPHA corporate network services (for example, email, web, Voice Over Internet Protocol [VOiP] telephony), or compromising a critical plant process (see **Table 3**, below). The red team automatically wins the competition if they completely halt the smartphone production line.

| Achievement | Point Value | Award Criterion | Validation Method |
|---|---|---|---|
| 1 | 5 points for each critical service stopped | Red team stops a Company Blue corporate critical service (for example, web, email, etc.) | White team will try to access the service. Points are awarded if the service is down for 10 minutes. |
| 2 | 10 points | Red team gains a remote shell on a Company Blue host | Red team shows the white team a screen capture video of a remote shell session and displays the IP address of the remote host. |
| 3 | 10 points | Red team has administrative privileges on a Company Blue host or network element | Red team executes a command on the host that requires administrative rights. Red team shows the white team a screen capture of the admin command executing. |
| 4 | 1 point for each IP address | Red team learns IP address of a host on the ICS network | Red team sends IP address to white team to validate. |
| 5 | Infinite | Company Blue's ICS process stops | White team witnesses the ICS process has stopped. |

**Table 3:** Red team point scoring system

## Recording war game activity

In this example, the mission of the white team is to record game play activity so that human actions can later be compared with predictions of computer security models, and the utility of the security models can be evaluated and hopefully improved. To accomplish a uniform sampling of actors, members of the red and blue teams will be chosen from academia, government institutions, or companies and team selection methods will be the same for both Event 1 and Event 2. As in a real scenario, blue team members can provide additional security measures during game play as needed. Red team members have a minimum amount of *a priori* technical information about the ALPHA system. Acquiring useful information for their attacks is part of the red team strategy.

Network traffic on the war game network will be recorded by the white team. The white team will also record strategic actions by red and blue team members, hopefully with assistance from the actual performers. After completing Events 1 and 2, the white team will present the game results at an open meeting and all teams will discuss lessons learned. Each team will have an opportunity to present its strategy and observations as well as recommendations for improving the war game. Point summaries will be discussed and awards will be given to the winning team. As mentioned, the purpose of having two events is to measure the effect of providing more in-depth training to the blue team for the second event.

**Post-game activities: comparison with models**
After the white team records game play activity, a computer security model will be used to analyse game play based on the initial system configuration and sample blue and red team strategies. This post-game activity is intended to test the validity of security models so that they can be improved for practical use as a security tool.

Next, a sample game-theoretic security model is described in which three games occur simultaneously at the physical layer, the cyber layer, and the management layer. This model assumes that all game players act rationally to optimize gain and an equilibrium strategy can be computed. For example, in the cyber game, the encounter between attacker and defender is described by a zero-sum game and rational moves of the two players are defined by saddle-points (Nash equilibria points) once the costs and awards of the two teams are known for all game strategies.

A sample cyber game (one of the three) is described next in which an attacker enters the attack surface and penetrates a series of layers guarded by the defender before arriving at the target (disabling the plant). The attacker devises a set $N_a$ strategies $\{s_{a,i}\} \in S_a$ where $S_a$ is the attack strategy space, where the attack strategies are identified by index $i$. Likewise, the defender has developed a set of $N_d$ strategies $\{s_{d,j}\} \in S_d$ where $S_d$ is the defence strategy space and $j$ is the defence strategy index. The sample model assumes that both attacker and defender have complete knowledge of the system and can consequently determine each other's strategies. The defender strategies are accomplished by selecting specific subsets of cyber-defence mitigations $\{m_{d,k}\} \in M_d$ where $M_d$ is the set of all mitigations, and $k$ is the mitigation index. There are $N_l$ layers that the attacker needs to penetrate. These layers are identified by the index $l$.

There are costs for both the attacker and the defender for each specific strategy tuple $\{i,j\}$. That is, given an attack strategy $i$ and a defence strategy $j$, the attacker suffers a cost $C_{a,ij}$ to accomplish his goal and the defender spends a cost $C_{d,ij}$ to deploy his or her defence strategy.

Finally, given a strategy choice $\{i,j\}$, the attacker is assumed to penetrate layer $l$ with probability $p_l(s_{a,i}, s_{d,j})$, or $p_{l,ij}$ in shorthand notation. If the attacker penetrates all $N_l$ layers, he or she reaches the target $T$ and obtains a benefit $b$.

It is assumed that when the attacker gains benefit $b$, the defender loses an equivalent value of his or her assets, so that the defender strives to keep the relevant portion ($b$) of his or her assets.

The attacker and defender expecting utility $u_a$ depends on the benefit $b$, the probability of penetrating all layers $P_{T,ij} = \prod_{l=1}^{N_l} p_{l,ij}$, and expended costs $C_{a,ij}$. For the attacker:

(1a) $\quad u_a(s_{a,i}, s_{d,j}) = b \prod_{l=1}^{N_l} p_l(s_{a,i}, s_{d,j}) - C_a(s_{a,i}, s_{d,j})$, or, in shorthand notation

(1b) $\quad u_{a,ij} = b\, P_{T,ij} - C_{a,ij}$

Likewise, for the defender:

(1c) $\quad u_d(s_{a,i}, s_{d,j}) = b[1 - \prod_{l=1}^{N_l} p_l(s_{a,i}, s_{d,j})] - C_d(s_{a,i}, s_{d,j})$, or, in shorthand notation

(1d) $\quad u_{d,ij} = b\,(1 - P_{T,ij}) - C_{d,ij}$

One game equilibrium can be calculated using a Stackelberg model in which the defender is the leader and the attacker is the follower. To choose his or her initial strategy $s_d^*$, the defender will seek to minimize the damages to his or her assets and the costs of defending his or her assets using an affordable defender strategy, which is equivalent to maximizing his or her utility $u_{d,ij}$. That is:

(2a) $\quad s_d^* = \underset{s_{a,i} \in S_a \; s_{d,j} \in S_d'}{\mathrm{argmax}}\; u_d(s_{a,i}, s_{d,j})$,

where $S_d' \subset S_d$ is the subset of affordable defender strategies.

After the defender chooses strategy $s_d^*$, the attacker selects his attack strategy $s_a^*$, that maximizes his utility $u_{a,ij}$:

(2b) $\quad s_a^* = \underset{s_{a,i}}{\mathrm{argmax}}\; u_a(s_{a,i}, s_d^*)$

The defender can then assess the probability $P_T^*(s_{a,i}, s_{d,j})$ that the attack reaches the target T by calculating

(2c) $\quad P_T^* = \prod_{l=1}^{N_l} p_l(s_a^*, s_d^*)$.

A second Stackelberg equilibrium can also be calculated by allowing the attacker to move first, followed by the defender.

While these equilibrium points are interesting mathematical entities, the simple game-theoretical model here makes many assumptions that were not necessarily true in a real war game of human teams of defenders and attackers in our December 2016 "Terra" cyber-physical war game held at ARL (Colbert *et al.* 2017). Some important differences between our simple model and a more realistic attack scenario are:

- The attacker entity may be a group of people that work 8-hour shifts around the clock, not a single person with a single utility function;

- The attacker entity may be a group of people that has a leader that manages people that execute the attacks. The group will likely have members that do not take direction well from the manager;
- The attacker (and/or defender) may not know how to navigate his or her strategy space because of unknown information (rules) at the beginning of the game;
- As information is learnt by both teams, the game rules and utilities may change;
- Similarly, system vulnerabilities and, consequently, game rules may be dynamic;
- The actors on the defence and attacking side may not make decisions according to a known utility function; for example, a nation state may take actions independent of the costs, and in a war game a red-team member may be more interested in displaying his or her prowess rather than helping his or her team;
- Zero-sum is an interesting concept for finding solutions, but is not realistic;
- Government politics or polices may change during a nation-state attack, which effectively modifies the rules of the game;
- Attackers and defenders do not always act rationally. In a group setting, the attacker entity's decisions, effectively made by a number of people, may be chaotic or very difficult to model easily with a probabilistic model such as instance-based learning (Gonzalez, Lerch & Lebiere 2003); and,
- As a real-life security situation unfolds, a series of (illogical) moves are made on each side as the (dynamic) game evolves. If the uncertainty of each move is too large, the solution (for example, a saddle point in strategy space) might not be useful to a system owner.

These issues should be considered and accommodated at some reasonable level before game-theoretical models can be developed into good attacker-defender security models for war game or real actors.

## Conclusion

This sample cyber framework describes a simple game-theoretic approach in which both attacker and defender have complete knowledge of the system, can infer all strategies of their opponents, and act in a rational manner toward maximizing their utility (minimizing their costs). Variations on this cyber model will be needed to accommodate deviations from ideal (rational) actors.

As mentioned earlier, SCADA and ICS systems are not merely a cyber network. They are affected by the state of the physical system attached to the network. Attacks focused on the physical system can penetrate the cyber network. As physical system changes occur on a much faster timescale than the cyber game, it can be modelled as a differential game and solved simultaneously with the cyber game to create a two-game security model, (Zhu & Basar 2015). In addition, the plant operator and the supervisor exist in an important and distinct management layer and their actions strongly influence both the cyber and physical layers. The policies and procedures for plant operations also play a part in the management layer game. The plant operator monitors critical elements of the plant processes and makes optimal choices—within given policy constraints—to maintain system operability.

To best accommodate influences from these three layers, a three-game model is proposed in which defender and attacker play in the cyber regime, physical control devices and perturbations (intentional or accidental) play in the physical regime, and operator and plant management play in an abstracted management layer. All three regimes and all players can affect each other in this complex, three-game model.

The amount of information revealed to the players is an important determinant in the outcome of the game. Assuming costs and behaviour can be modelled well, the attacker will often not have complete knowledge of the three regimes when they begin their attacks, as the security model may assume. In fact, in ARL's two war game events, the red team does not have complete knowledge of the system when it begins game play. It may have done some reconnaissance work, but will be missing important pieces of information when it initiates attack strategies. This lack of information will affect the team's path taken through attack space.

Clearly, the purpose of the post-game activities, and arguably, the war game itself, is to identify significant deficiencies in the security model so that it can be improved for practical use. By adding more complexity to the model and recalculating results, an improved SCADA security model can be developed and validated using recorded actions such as those from this sample SCADA cyber war game.